\documentclass[10pt, conference]{IEEEtran}

\usepackage[T1]{fontenc}
\usepackage{paralist}
\usepackage{cite}
\usepackage{amsmath,amssymb,amsfonts}
\usepackage{algorithmic}
\usepackage{graphicx}
\usepackage{textcomp}
\usepackage{xcolor}
\usepackage[inline]{enumitem} 
\usepackage{enumerate} 
\usepackage{multirow}
\usepackage{balance}
\usepackage{hyperref}
\usepackage[ruled,vlined,linesnumbered]{algorithm2e}
\usepackage{tablefootnote}
\usepackage{subcaption} 
\usepackage{dblfloatfix}

\SetCommentSty{mycommfont}

\usepackage{pifont}

\usepackage{xcolor}

\makeatletter
\IEEEtriggercmd{\reset@font\normalfont\footnotesize}
\makeatother
\IEEEtriggeratref{1}

\newcommand{\GGML}{\textit{GGML}\xspace}
\newcommand{\qkw}{\textit{Q3\_K}\xspace}
\newcommand{\qki}{\textit{Q8\_K}\xspace}
\newcommand{\qk}{\textit{Q3\_K}\xspace}

\newcommand{\qkernel}{\textit{MatMul\_Q3\_K\_Q8\_K}\xspace}

\newcommand{\LPP}{\textit{llama.cpp}\xspace}

\begin{document}
\bstctlcite{IEEEexample:BSTcontrol}

\title{Designing Efficient LLM Accelerators \\ for Edge Devices}


\author{Jude Haris, Rappy Saha, Wenhao Hu, Jos\'e Cano \\
\emph{School of Computing Science, University of Glasgow, Scotland, UK}
}


\maketitle


\begin{abstract}

The increase in open-source availability of Large Language Models (LLMs) has enabled users to deploy them on more and more resource-constrained edge devices to reduce reliance on network connections and provide more privacy.
However, the high computation and memory demands of LLMs make their execution on resource-constrained edge devices challenging and inefficient.
To address this issue, designing new and efficient edge accelerators for LLM inference is crucial. FPGA-based accelerators are ideal for LLM acceleration due to their reconfigurability, as they enable model-specific optimizations and higher performance per watt.
However, creating and integrating FPGA-based accelerators for LLMs (particularly on edge devices) has proven challenging, mainly due to the limited hardware design flows for LLMs in existing FPGA platforms.

To tackle this issue, in this paper we first propose a new design platform, named \textit{SECDA-LLM}, that utilizes the SECDA methodology to streamline the process of designing, integrating, and deploying efficient FPGA-based LLM accelerators for the \textit{llama.cpp} inference framework.
We then demonstrate, through a case study, the potential benefits of SECDA-LLM by creating a new MatMul accelerator that supports block floating point quantized operations for LLMs.
Our initial accelerator design, deployed on the PYNQ-Z1 board, reduces latency ($1.7$ seconds per token or $\sim2$ seconds per word) by $11\times$ over the dual-core Arm NEON-based CPU execution for the TinyLlama model.






\end{abstract}


\section{Introduction}

Large language models (LLMs) are an emerging class of machine learning (ML) systems geared toward learning from huge text-based datasets.
LLMs such as GPT-3\cite{brownLanguageModelsAre2020} have revolutionized the ability of Artificial Integillence (AI) systems to understand and generate human language.
Due to innovative changes in model architecture, training methods, and through the help of the popularity of online services like ChatGPT~\cite{rayChatGPTComprehensiveReview2023}, the field of LLMs is evolving rapidly.

The number of everyday users is also growing rapidly due to the myriad of use cases from translation~\cite{yaoBenchmarkingLLMbasedMachine2024}, classification~\cite{sunTextClassificationLarge2023}, code generation~\cite{liuYourCodeGenerated2023} to healthcare~\cite{clusmannFutureLandscapeLarge2023}.
Additionally, cloud-based LLM services are currently the go-to method of access to LLMs for everyday users, but as the availability of open-source LLMs and datasets increases, especially over the last few years, the need for edge-based, localized access and execution of LLMs has become more sought after.
Massive community-driven pushes have facilitated easy access to LLMs and rapid prototyping of new models and optimizations to enable efficient LLM inference on edge devices.
At the forefront of these pushes is the GPT-Generated Model Language~\cite{GGML} (GGML).
GGML is a tensor library for ML specialized in enabling large models and high performance on commodity hardware.
Furthermore, the \GGML's \LPP~project~\cite{LLP} is specialized towards running LLMs on edge devices, supporting LLM inference on commodity CPUs and GPUs.

Unfortunately, LLMs can be very computationally demanding, even for inference.
In addition, due to their large memory footprint, they require high memory capacity and bandwidth.
These properties of LLMs make them challenging to execute on resource-constrained edge devices.
For example, running LLMs on mobile phones or Internet-of-Things devices (IoT) devices is in some cases impossible due to memory constraints.
Hence, there is a great demand for developing and deploying custom hardware accelerators to efficiently run these LLMs on resource-constrained edge devices.
Fortunately, Field Programmable Gate Arrays (FPGAs) are ideal for designing new flexible and power-efficient accelerators that can take advantage of LLM optimizations, such as block floating point quantization.
While some FPGA-based accelerators~\cite{Khan2021NPE,Lu2020Transformer} already exist for LLM inference at the edge, with constant changes to LLM architectures and optimizations we are in need of new specialized FPGA-based accelerators.

To create new and innovative FPGA-based accelerator architectures for LLM inference at the edge, we need ways to quickly prototype and evaluate LLM-based inference accelerators to reduce development costs and increase design space exploration.
Hence, we propose SECDA-LLM, a new platform for designing, integrating and deploying specialized accelerators for LLMs at the edge.
SECDA-LLM employs the SECDA design methodology~\cite{harisSECDAEfficientHardware2021}, and similar to SECDA-TFLite~\cite{harisSECDATFLiteToolkitEfficient2023a}, it provides the user with the ability to quickly prototype accelerator designs with the target application framework, in this case, \LPP~project.
Our SECDA-LLM platform enables the designer to consider hardware-software co-design optimizations in terms of both algorithmic and hardware implementations~\cite{gibson_dlas_2024}, and makes deployment of LLMs through FPGA-based accelerators effortless.
The contributions of this work are as follows:

\begin{itemize}
    \item \textit{SECDA-LLM}, a new design platform using the SECDA methodology which enables the design, integration and deployment of FPGA-based accelerators for LLMs on resource-constrained edge devices.

    \item A case study to demonstrate \textit{SECDA-LLM}, where we prototype and deploy a new accelerator to efficiently execute quantized MatMul operations during LLM inference.

    \item Evaluation of our initial accelerator design executing the TinyLlama model~\cite{zhangTinyLlamaOpenSourceSmall2024} on the PYNQ-Z1~\cite{pynqz1} board, where we achieve a $11\times$ speedup over dual-core ARM NEON-based CPU execution.
    
        
\end{itemize}




\section{Background and Related Work}



\subsection{Large Language Models }

LLMs are a family of ML models that use the Transformer~\cite{vaswani2017attention} architecture as the key component, and are pre-trained on large amounts of language data.
People usually use them by fine-tuning with downstream task-specific datasets.
LLMs usually have large amounts of parameters.
For example, Llama is designed to start with 7B parameters~\cite{touvron2023llama}.
Also, many types of language-generating LLMs auto-regressively compute the next tokens (chunks of text) by using the previously cached information.
This computation paradigm, called KV cache~\cite{kwon2023efficient}, introduces a large amount of memory overheads while improving performance.

Quantization techniques are extensively employed to deploy parameter-intensive LLMs on resource-constrained edge devices.
For example, utilizing 8-bit quantization facilitates the retention of model accuracy while significantly reducing the model size~\cite{Zafrir2019Q8BERT, Wan2024ASP-DAC}.
In addition, prior studies have demonstrated efforts in experimenting with smaller number of bits, such as 4-bit quantization, to enable the operation of LLMs on edge devices while upholding accuracy levels~\cite{Shen2024EdgeQAT}. 
Another technique that can be considered is block floating point (BFP) quantization, and there have been some works~\cite{Rouhani2023Microscaling} comparing the efficacy of BPF to traditional integer (8/4-bit) quantization.



\subsection{Inference framework: \LPP}

\LPP~\cite{LLP} is a pure C/C++ library with minimal external dependencies for enabling LLMs inference on a wide range of hardware.
Currently, \LPP supports a wide range of LLMs, including some multi-modal and custom-defined models.
Additionally, it supports 1.5-bit, 2-bit, 3-bit, 4-bit, 5-bit, 6-bit, and 8-bit BFP quantization.
In \LPP, BFP quantization is leveraged to quantize the weights of the LLMs, and it includes few quantization variations.
These variations are typically denoted as \( Q{x\_y} \), where \( x \) represents the number of bits per weight and \( y \) denotes the type of quantization.

\LPP~\cite{LLP} employs the GPT-Generated Unified Format model format (GGUF) to represent LLMs.
Within this format, it is possible to represent the weights of an LLM with as few as 1.5 bits using BFP quantization.
These quantized weights enable users to run LLMs on resource-constrained edge devices such as the Raspberry Pi and the Pixel phone~\cite{simonwillison}.

LLM inference is possible on edge devices equipped with CPUs or GPUs due to the optimized support provided by \LPP~\cite{LLP} for AVX, AVX2, and AVX-512 on x86 architectures as well as custom CUDA kernels for running on NVIDIA GPUs.
However, LLM inference on resource-constrained devices, especially on FPGAs is not straightforward, as the design process for new FPGA-based accelerators has not been integrated with inference platforms like \LPP yet.


\subsection{FPGA-based acceleration for LLMs}

Previous works have described acceleration ideas for running LLMs on FPGAs~\cite{Kachris2024Survey}.
The focus was generally to accelerate LLM inference.
Additionally, they have addressed the challenge of executing a wide variation of LLMs on FPGAs by proposing an overlay FPGA-based processor~\cite{Khan2021NPE}.
However, research has yet to focus on developing design flows and enabling explorations of new FPGA-based acceleration ideas for the latest LLMs.
While transformer accelerator design that commences with the C/C++ code of an LLM exists~\cite{Lu2020Transformer}, their primary focus lies on the accelerator design aspect, not necessarily emphasizing the design process itself, providing no design methodology (e.g., no hardware simulation for prototyping) for accelerator development.

\begin{figure} [t]
 \centering
 \includegraphics[width=0.9\columnwidth]{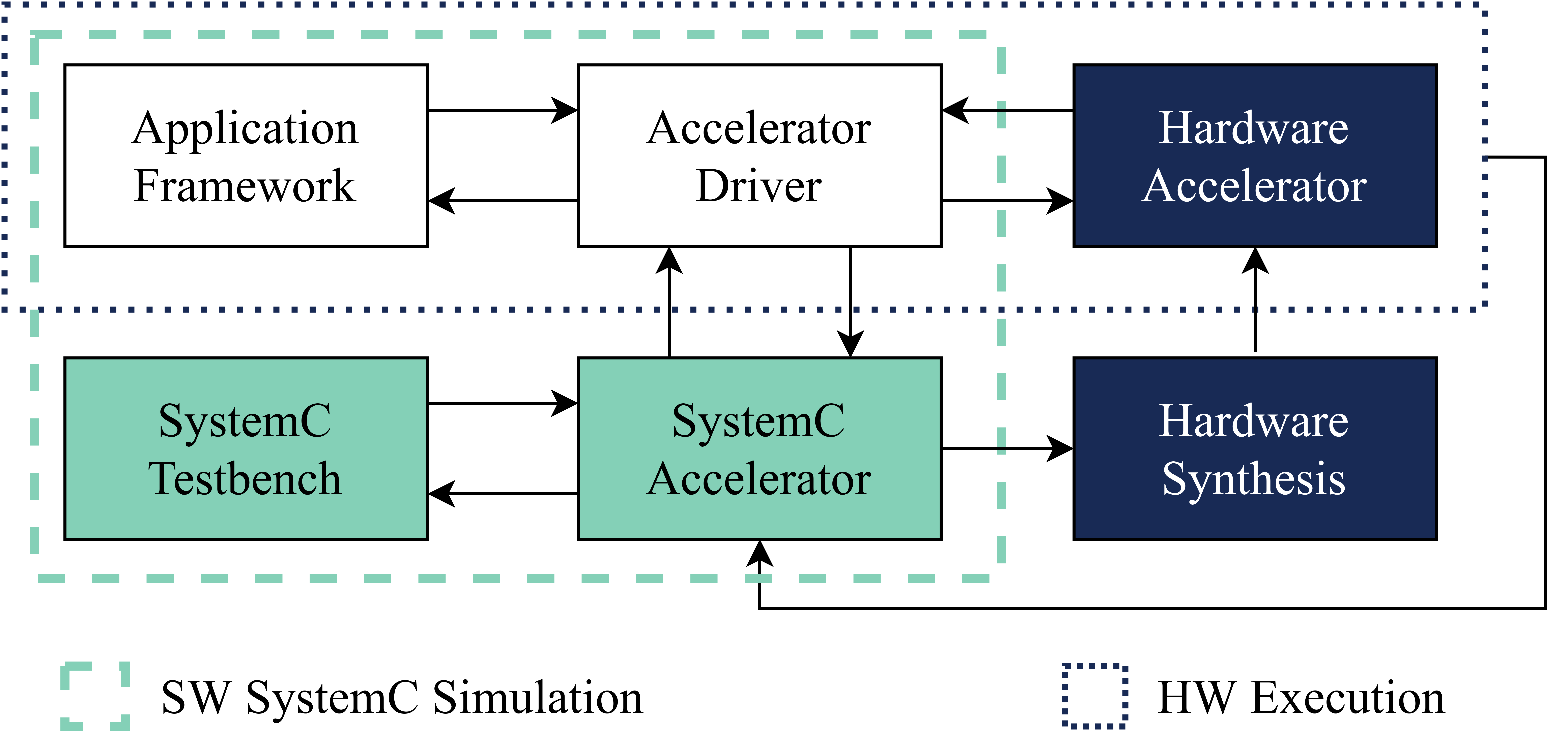}
  \caption{\label{fig:secda_method} Overview of the SECDA methodology~\cite{harisSECDAEfficientHardware2021}.
  Components in the dashed lines correspond to simulation, and in the dotted lines to execution on real hardware.}
\end{figure}

As such, we aimed to employ SECDA (SystemC Enabled Co-design of DNN Accelerators)~\ref{fig:secda_method}, a hardware-software design methodology to efficiently produce optimized inference accelerators for edge devices using FPGAs.
SECDA uses SystemC~\cite{IEEE2012systemc} as an accelerator simulation framework, allowing candidate designs to be efficiently iterated upon.
Additionally, SECDA uses SystemC High-Level Synthesis (HLS) to produce a synthesizable design based on the same SystemC accelerator definition.
One key aspect of SECDA is the full integration of the design process with the target application framework.
For LLMs inference, \LPP is the ideal target application framework.
With the integration of \LPP as the application framework, it becomes feasible to convert an LLM to the GGUF format and execute LLMs on edge FPGA-based platforms.







\section{SECDA-LLM}

\begin{figure}[!ht]
 \centering
 \includegraphics[width=0.9\columnwidth]{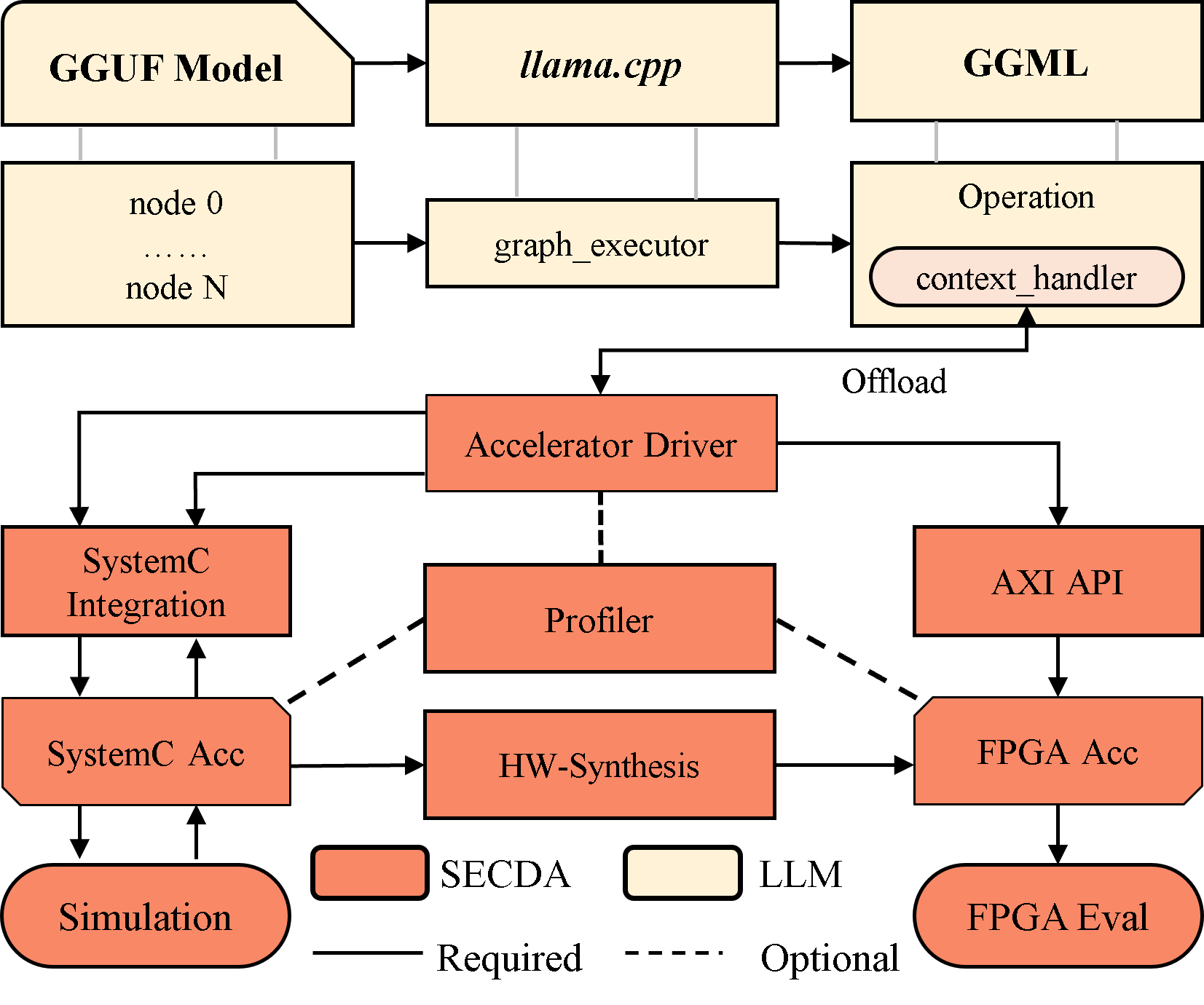}
  \caption{\label{fig:secda_llm} Overview of SECDA-LLM. Key SECDA components are highlighted in orange, and the LLM components are highlighted in beige.}
\end{figure}

SECDA-LLM is a specialized platform for creating FPGA-based LLM accelerators for edge devices using the SECDA methodology~\cite{harisSECDAEfficientHardware2021} within the \LPP environment.
Figure~\ref{fig:secda_llm} outlines the main components of SECDA-LLM.
The platform simplifies the accelerator design process by integrating the SECDA tools (e.g., AXI-API, profiler) , thus allowing a seamless connection between the SECDA design environment and the target application framework, \LPP.
This integration enables developers to begin prototyping and integrating their new accelerator designs with minimal setup costs.

The rest of this section provides details about SECDA-LLM at:
\begin{enumerate*}[label=\roman*)]
    \item how it is integrated with \LPP;
    \item how it enables the accelerator designer to prototype and simulate new designs using SystemC~\cite{IEEE2012systemc};
    \item the ease of hardware evaluation;
    \item the profiling and performance analysis capabilities.
\end{enumerate*}


\subsection{Integration with \LPP}

Figure~\ref{fig:secda_llm} shows that SECDA-LLM builds upon the core \LPP project.
Our current integration is through \LPP's `main' example project, which enables users to run LLMs with minimal overhead.
We can connect into \LPP once it calls any of the \GGML's operations.

Depending on our target operation(s), we create additional connection points from the \GGML library to the SECDA environment.
During these connections, we ensure the creation of a context handler to pass from the \GGML environment to the SECDA environment; the context handler includes memory pointers, memory-mapped model data, access to relevant inputs tensors, quantization, and layer parameters.


\subsection{SECDA Environment}

Within the SECDA components shown in Figure~\ref{fig:secda_llm}, the accelerator designer can start quickly prototyping the initial accelerator design and driver code.
First, the user is required to create the initial driver, a simple C++ class that will gain access to the context handler provided by the offload call from~\GGML.
Second, the developer must create an initial SystemC description of their accelerator.
Then, the user can instantiate their desired data communication channels between the driver and accelerator using data interfaces provided within the SECDA environment (e.g, AXI4-S, AXI-MM and AXI-Lite).
The developer can use these data channels for SystemC end-to-end simulation.


\subsection{SystemC Simulation}

SystemC end-to-end simulation is a crucial step in the SECDA methodology; therefore, SECDA-LLM provides access to SystemC simulation.
The simulation-based design loop is shown at the bottom left half of Figure~\ref{fig:secda_llm}.
Once the driver and accelerator are connected through the desired data communication channels, the user can perform end-to-end simulations of LLMs using SECDA-LLM.
With simulation enabled, the designer can quickly prototype new driver and accelerator features, verifying correctness, profiling performance and modeling control flow behavior within their design.
Therefore, the hardware developer is able to rapidly iterate through their design process, through end-to-end simulation, to meet the target performance.


\subsection{Hardware Evaluation}

With simulation-based evaluation, the designer can quickly make fast, broad design changes.
Once satisfied with a given design, the designer can quickly take the SystemC-defined design and perform High-level synthesis (HLS) and logic synthesis (HLX) through the hardware synthesis tool provided by SECDA-LLM to map it to the target FPGA, as shown on at bottom right of Figure~\ref{fig:secda_llm}.
Additionally, as SECDA-LLM is integrated with the \LPP project, we can leverage the \LPP project's compilation flow to generate pre-defined applications that use the LLMs through the \LPP's interface.
These generated applications will now have complete access to the driver and accelerator for execution on an FPGA-enabled device; see Section~\ref{subsubsec:exp_setup} for details.

A major benefit of the SECDA methodology, and therefore SECDA-LLM, is that we can reuse the driver and accelerator completely.
Therefore, for actual FPGA evaluation the designer does not need to make any changes to the driver to enable real hardware execution, as the SECDA data interfaces switch between simulation and FPGA execution through a simple \textit{"SYSC" }compiler flag.
Once the accelerator is mapped to the target FPGA, the designer can evaluate its performance with their target applications.


\subsection{Profiling}


Through SECDA-LLM, we provide two types of profiling: simulation profiling and execution time profiling.
The \emph{profiler} module shown in Figure~\ref{fig:secda_llm} highlights how the profiling interacts with both the accelerator design and driver.
Additionally, we are able to leverage any additional profiling tools provided by the \LPP project.

\subsubsection{Simulation profiling}

End-to-end SystemC simulation can be used to quickly evaluate the potential performance impact of changes to the hardware and software components of the accelerator design and verify the correctness of the implementation.
To profile the end-to-end simulation, the developer typically needs to add additional profiling code to keep track of hardware and software metrics throughout the end-to-end LLM inference.
The \emph{profiler} module provided within SECDA-LLM enables the quick and easy method to set up capture points to profile from the accelerator.
The capture points can record different metrics of the accelerator and hardware submodules.
Metrics include clock cycle counts and the dynamic utilization of processing elements and accelerator buffers.

\subsubsection{Execution profiling}

During the hardware evaluation, SECDA profiling provides execution time for the custom driver and accelerator.
This type of profiling helps the designer understand the performance bottlenecks caused by driver-accelerator interactions.
For instance, a designer may opt to profile time spent:
\begin{enumerate*}[label=\roman*)]
    \item Sending input data;
    \item Waiting for the accelerator to execute operations;
    \item Unpacking output data received from the accelerator.
\end{enumerate*}
The analysis of these detailed execution time breakdowns can motivate both accelerator and driver design choices.
Additionally, execution profiling of an FPGA evaluation run can be used to profile driver execution times, which can be combined with SystemC-reported simulation times for the accelerator.
This would estimate end-to-end execution time in terms of both CPU and accelerator.



\section{Case study}

To demonstrate our SECDA-LLM platform and how it provides a quick and efficient design flow for developing LLM accelerators for edge devices, we develop a new custom FPGA-based accelerator for BFP quantized LLM inference.



\subsection{Target Problem}

We target the acceleration of MatMul operations within our target LLM, as MatMul represents about $97\%$ of the computations.
Specifically, we accelerate the GGML's \qkernel kernel, which uses 3-bit weights and 8-bitinputs with BFP quantization.


Both weights and inputs are stored in what is called "super-blocks" (SBs); these SBs are critical in maintaining LLM accuracy by adjusting mathematical scaling during computation.
With the \qkw format used for weights, each SB can represent 256 weights ($N_w$), where the SB is partitioned into 16 tiles ($N_{tiles}$) and each tile contains a scaling factor (6-bits) and 16 weights (3-bits). Additionally, each SB has one super-scaling factor (16-bits), which equates to $\sim$3.5 bits-per-weight.
With the \qki format, used for inputs, each SB contains 256 inputs (8-bits) and a single super-scaling factor \textit{SSF} (16-bits), which equates to 8$\sim$ bits-per-input.




\subsection{Accelerator Design}

\begin{figure}[!t]
 \centering
 \includegraphics[width=0.95\columnwidth]{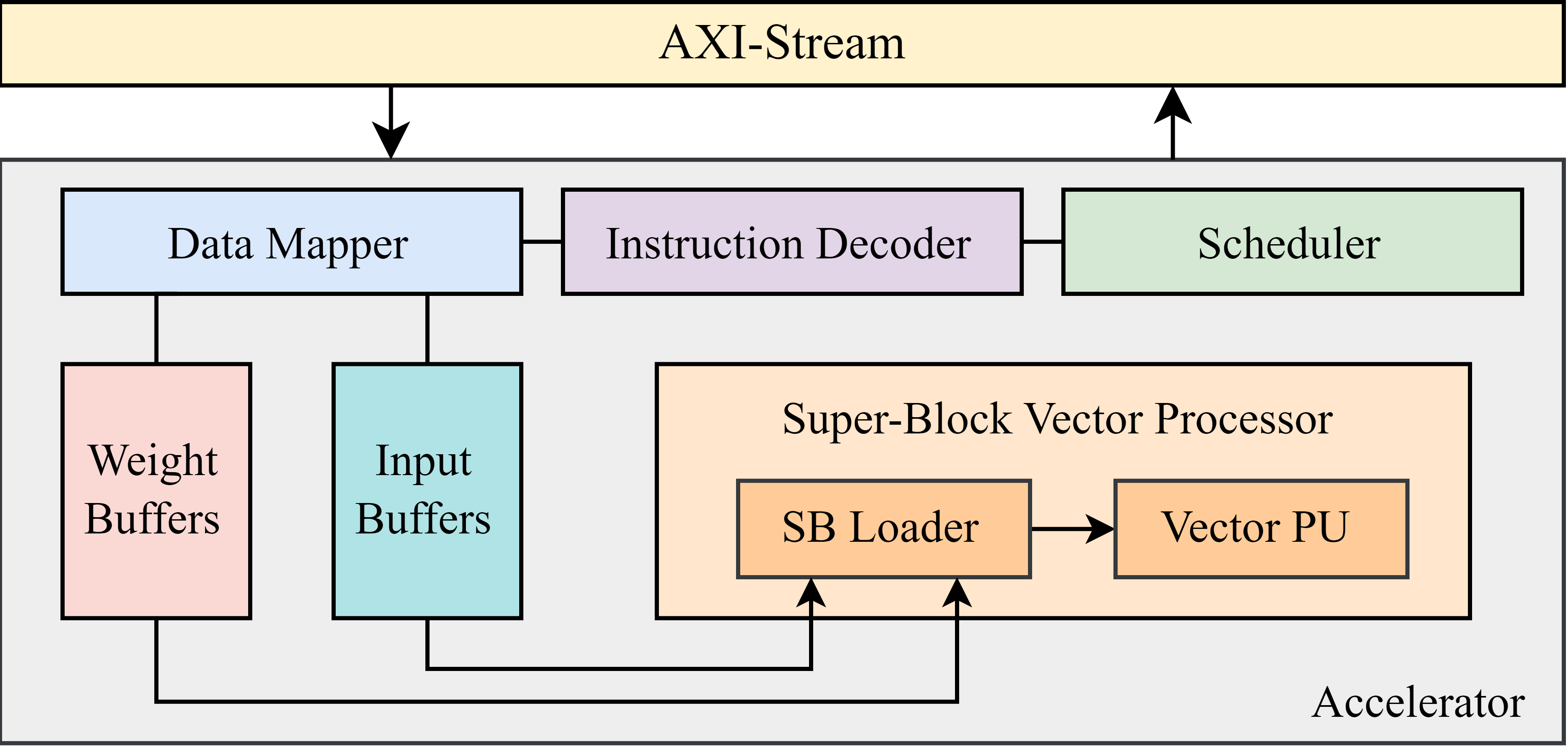} 
        \caption{\label{fig:llm_acc} Overview of our block floating point quantized accelerator design for GGML's \qkernel kernel.}
\end{figure}

Our accelerator design, shown in Figure~\ref{fig:llm_acc}, contains an \emph{instruction decoder}, a \emph{data mapper}, a \emph{scheduler} and the \emph{Super-Block Vector Processor (SBVP)}:

\begin{itemize}
    \item The \emph{instruction decoder} loads and decodes instructions from the AXI-Stream and then communicates the instruction throughout the rest of the accelerator.
    
    \item The \emph{data mapper} parses the incoming data stream and maps the weight and input super-blocks into their respective weight and input buffers. 
    Our mapping scheme enables efficient data access, so that the \emph{SBVP} can compute without stalling the computation pipeline.
    
    \item The \emph{SBVP} efficiently computes the dot product between the SB of weights and inputs while scaling the computation according to the SB scaling factors.

    \item The \emph{scheduler} tiles the MatMul problem according to the dimension of the target layer. 
    Additionally, it synchronizes and accumulates the output data produced by the SBVP and sends the results back to the main memory using the AXI-Stream.
\end{itemize}


\subsection{Evaluation}
\label{subsubsec:exp_setup}


We evaluate our accelerator design on the PYNQ-Z1 board~\cite{pynqz1}, which contains a Xilinx Z020 edge FPGA and a 650MHz dual-core Arm Cortex-A9 CPU.
We execute inference for the TinyLlama model~\cite{zhangTinyLlamaOpenSourceSmall2024}, that contains 1.1B parameters (460 MB), trained on the Guanaco dataset~\cite{joséphus_cheung_2023}.
This model contains various BFP quantization levels, but most layers are quantized to \qk.
Note that with~\LPP you can apply different levels of quantization to reduce model size as required.


For our experiments, we use the \LPP project's `main' program cross-compiled for our target CPU architecture, ARMv7a, with Neon vector instructions enabled alongside our accelerator driver.
We execute the TinyLlama model utilizing our FPGA-mapped accelerator to offload the \qkernel  layers, to obtain an initial speed of $1.7$ seconds per token ($\sim2$ seconds per word).
This provides a $11\times$ speedup over CPU-only inference, drastically improving the usability of LLMs on such a resource-constraint device. 
Therefore, with SECDA-LLM we could quickly implement, integrate and evaluate our new accelerator design.

\section{Conclusion}

We introduced SECDA-LLM, a novel platform that simplifies the creation of specialized hardware accelerators for LLM inference on resource-constrained edge devices. 
SECDA-LLM integrates the SECDA methodology within \LPP, enabling developers to access the SECDA tools (e.g., AXI-API, profiler), which can be used to effectively co-design new FPGA-based accelerators for LLMs with ease.
As a case study, we presented a quantized MatMul accelerator design that optimizes LLM inference (by $11\times$ over the dual-core Arm NEON-based CPU execution) for the TinyLlama model.
Future work will expand SECDA-LLM into an open-source platform for collaborative development and continuous improvement of LLMs' performance on resource-constrained edge devices.

\section*{Acknowledgment}

This work was partially supported by the Engineering and Physical Sciences Research Council (grant EP/R513222/1) and the EU Project dAIEDGE (GA Nr 101120726).



\balance

\bibliographystyle{IEEEtran}
\bibliography{bib}

\end{document}